\documentstyle[aps,prb,epsfig,floats]{revtex}

\newcommand{\unite}[1]{\mbox{$\rm #1$}}

\newcommand{\be}{\begin{equation}}
\newcommand{\ee}{\end{equation}}

\begin{document}

\draft

\preprint{}

\twocolumn[\hsize\textwidth\columnwidth\hsize\csname %
@twocolumnfalse\endcsname

\title{Excess Spin and the Dynamics of Antiferromagnetic Ferritin}

\author{J.G.E.\ Harris, J.E.\ Grimaldi, and D.D.\ Awschalom}

\address{Department of Physics, University of California, Santa 
Barbara, CA 93106}

\author{A.\ Chiolero and D.\ Loss}

\address{Department of Physics and Astronomy, University of Basel,\\
Klingelbergstrasse 82, 4056 Basel, Switzerland}

\date{\today}

\maketitle

\begin{abstract}

Temperature-dependent magnetization measurements on a series of
synthetic ferritin proteins containing from 100 to 3000 Fe(III) ions
are used to determine the uncompensated moment of these
antiferromagnetic particles. The results are compared with
recent theories of macroscopic quantum coherence which explicitly
include the effect of this excess moment. The scaling of the excess
moment with protein size is consistent with a simple model of finite
size effects and sublattice noncompensation.

\end{abstract}

]
% 73.40.Gk: Tunneling
% 75.60.Jp: Fine-particle systems
% 75.10.Jm: Quantized spin model
% 75.30.Gw: Magnetic anisotropy
% 03.65.Sq: Semi-classical theories and applications
% 76.30.Fc: Iron group (3d) and impurities (Ti-Cu)

\narrowtext

In nanometer-scale magnetic particles it is possible to observe a number of
phenomena which do not exist in bulk samples.  The evolution of
magnetic order from individual atoms to large
clusters,\cite{gidreview} thermal relaxation of the magnetization from
a metastable state,\cite{wernsdorfer} the different roles of surface
and interior atoms,\cite{kodama} and the quantum mechanical dynamics
of the order parameter\cite{MQTferro} have all received considerable
experimental and theoretical attention in the last several years.
Here we present measurements on a series of
well-characterized samples of biomimetic antiferromagnetic particles
with sizes from 100 to 3000 Fe(III) ions per particle.  By
separating the bulk from the surface contributions to the
magnetization, we explore the connection between two of these
phenomena: the role of the excess moment in the macroscopic quantum
coherence of antiferromagnetic particles.

Past theoretical work suggested that the N\'eel vector of a small 
antiferromagnetic particle could exhibit macroscopic quantum coherence 
(MQC), in which it tunnels resonantly between degenerate easy 
directions, at a rate accessible to experiment.  
\cite{intro:Barb90,intro:Krive90} Measurements of the magnetic noise 
spectrum and ac susceptibility of the antiferromagnetic cores of the 
protein ferritin revealed a resonance whose frequency scaled with 
particle size, applied magnetic field, temperature, and interparticle 
separation in qualitative agreement with theoretical predictions.\cite 
{intro:Awsch92,intro:Gid94,intro:Gid95} In the interpretation of these 
results, it was assumed that any excess moment of the ferritin cores 
would follow the dynamics of the N\'eel vector without affecting it.  
Measurements on natural ferritin have shown that the cores do have a 
small net magnetic moment ($\approx$100's of 
$\mu_{B}$),\cite{intro:Awsch92,Blaise67,mohie,applic:Kil95,makhlouf} 
presumably due to the preferential population of one magnetic 
sublattice during the formation of the particles.  Recent theoretical 
work\cite{model:Loss92,Chio97} predicts that an excess spin 
$\approx$100 will have a small but appreciable effect on the MQC 
frequency of an antiferromagnetic particle.  Motivated by this 
prediction and the opportunity to test long standing 
models\cite{Neelscaling} of the size dependence of the excess moment 
in small antiferromagnetic particles, we have measured the excess 
moments of several artificially synthesized ferritin samples.

Occurring in a wide range of plants, animals, and bacteria, ferritin 
consists of an organic hollow spherical shell with an inner core 
$\sim$80 \AA\ in diameter.  This protein absorbs Fe ions through 
channels in its shell where they nucleate into an insulating 
antiferromagnetic crystal ($T_N=240 K$\cite{applic:Baum89}) similar to 
ferrihydrite.  Natural ferritin can contain at most 4500 Fe(III) ions, 
and typically contains an average of 2000.  Using synthetic chemical 
techniques, it is possible to prepare samples in which each protein 
shell contains a fairly well-specified number of Fe(III) ions.  The 
proteins used in these measurements have been extensively 
characterized elsewhere.\cite{intro:Gid94,intro:Gid95,JAP:Gider96} The 
samples have nominal loadings of $n$=100, 250, 500, 1000, 2000, and 
3000 Fe(III) ions (ionic moment $\mu_{Fe(III)}=5.92 \mu_{B}$) per 
protein.  Transmission electron microscopy (TEM) measurements of the 
mean and variance of the core diameters have been made for all but the 
smallest two samples, and published elsewhere.\cite{intro:Gid95} For 
magnetization measurements, a dilute ($\sim$0.5mg/ml) solution of a 
sample is dried on a polypropelene film and mounted on a twisted Cu 
wire in a commercial SQUID magnetometer.  Measurements are made at 
temperatures ranging from T=4-300K and applied fields H=0-5 T.

Typical M(H) curves for the $n$=2000 and $n$=3000 sample are shown in 
Fig.1 and reveal the presence of two components: one which saturates 
at large H, and a second that is approximately linear in 
H.  N\'eel\cite{Neelmag} modeled small antiferromagnetic 
particles with net moments as an order parameter 
(essentially the N\'eel vector) possessing a magnetic moment $\mu$ as 
well as parallel and perpendicular susceptibilities $\chi_\|$ and 
$\chi_\perp$.  The energy of such a particle can be written as

\be
E = -{1\over2}\chi_\| H^{2} \cos^{2}\psi - {1\over2}\chi_\perp
H^{2} \sin^{2}\psi - \mu H \cos\psi
\label{modelE}
\ee
where $H$ is the applied field and $\psi$ is the angle between the
N\'eel vector and $H$.  The thermodynamic magnetization $M(H)$ =
${k_{B}T}{\partial\over{\partial H}}$ $\ln[Z]$, where $Z$ is the partition function,
can be calculated explicitly, but for $H < \mu/\chi_\perp$ (which
holds for all measurements here), we use the approximate
expression:\cite{Neelmag,Savthesis}
\be
M(H) = \chi_\| H + (\mu + 2(\chi_\perp - \chi_\|)k_B T/\mu)L[{\mu
H\over k_{B}T}] ,
\label{modelM}
\ee
where $L$[x] is the Langevin function.  Because 
Eq.(\ref{modelM}) assumes thermodynamic equilibrium, all of our 
measurements are made above the blocking temperature $T_{B}$ (which is 
size dependent\cite{intro:Gid95}) of each sample.  This also justifies 
our ignoring anisotropy energy terms in Eq.(\ref{modelE}).

Figs.1(a) and 1(c) show a fit to Eq.(\ref{modelM}) for the data
from the $n$=2000 and $n$=3000 samples a few Kelvin above
$T_{B}$.  The fit fails in the region where $M(H)$ has the most
curvature, as there is a non-negligible spread in the
particle sizes in each sample. This results in a departure from
Eq.(\ref{modelM}) which is difficult to model without knowing the precise
size distribution in the sample.  However, we can exploit the fact
that the susceptibility near zero field is simply the {\it average}
susceptibility of the sample.  We first subtract the component of M(H)
which is linear at high fields where the Langevin function has
saturated to remove both the contribution from the
antiferromagnetic bulk (the first term in Eq.(\ref{modelM})) and any
diamagnetic background.  The data is then normalized so that the
saturation moment is unity and $\mu$ is extracted from the slope of
the low field
data, ${\mu/(3k_{B}T)}$.  This approach
has the advantage of being insensitive to the total number of
particles in the sample as well as to the details of its size
distribution since it returns the {\it average} $\mu$ of the sample.
Figs.1(b) and 1(d) show fits to the high- and low-field data from
Figs.1(a) and 1(c) using Eq.(\ref{modelM}).  The fits represent the
theoretical M(H) of a particle with $\mu$ equal to the average $\mu$
of the sample.

This procedure can be used at any temperature above $T_{B}$ (i.e., 
whenever the magnetic moments are in thermal equilibrium on the time 
scale of the measurement).  We repeated measurements of the type shown 
in Fig.1 over a temperature range from roughly $T_{B}$ to 4$T_{B}$.  
In the samples with larger cores (Fig.2(a)), we find a weak 
temperature dependence of the extracted $\mu$, increasing by 
$\sim$20\% over a factor of 4 in temperature.  For the smallest cores 
(Fig.2(b)), the temperature dependence of $\mu$ is quite strong, 
increasing by roughly a factor of two over the same range.  A similar 
trend has been observed in natural ferritin as well as NiO 
particles.\cite{applic:Kil95,makhlouf} The reason for this temperature 
dependence is not clear.  It should be noted that the model behind 
Eqs.(\ref{modelE}) and (\ref{modelM}) does not take into account any of 
the microscopic phenomena which might alter the properties of a small 
antiferromagnetic particle except in as much as these effects can be 
modeled by $\mu$, $\chi_\|$ and $\chi_\perp$.  Weaker exchange,\cite 
{Neelscaling} strong radial anisotropies,\cite{rado} and 
frustration\cite{kodama2} can exist at the surfaces of such particles, 
and may be responsible for the observed temperature dependence of 
$\mu$.  Multiple sublattices can also exist in very small particles of 
a material which is antiferromagnetic in the bulk.\cite{kodama} At the 
low concentrations used here the typical interparticle dipolar fields 
should be well below 1 G, too small to account for these effects.  For 
comparison with MQC, which is only observed below 200 mK, we 
extrapolate the linear temperature dependence shown in Fig.2 to T=0 in 
order to extract the relevant excess moment.

The result of this extrapolation is shown in Fig.3, where $\mu$ is
plotted vs.  $n$.  The vertical error bars represent the combined
effects of the reproducibility between identically prepared
samples and the uncertainties in the linear extrapolations of Fig.2,
$\sim$10\%.  Because the extracted $\mu$ corresponds to the excess
moment averaged over the sample, the
horizontal error bars do not represent the variance in particle size,
but rather the uncertainty in the mean particle size, estimated from
the discrepancies between the nominal loading and the particle size
measured by TEM ($\sim$20\%).  This is probably an underestimate in
the case of the smallest two samples.  The data are fit by a 0.56 $\pm$
0.05 power law.  If one plots $\mu$ in units of
$\mu_{Fe(III)}$, the coefficient of the power law fit is 1.15, quite close
to unity.  N\'eel has suggested three models of imperfect sublattice
compensation in small antiferromagnetic particles.\cite{Neelscaling}
In the first, an antiferromagnetic particle has a surface
consisting of sites belonging to one sublattice only, giving $\mu =
c n^{2\over3} \mu_{Fe(III)}$.  The proportionality constant $c$ is
roughly 4 for the platonic solids.  In the second model, the surface
sites are distributed randomly between the two sublattices; then one
has a random walk over the surface, and so $\mu = c^{1\over2}
n^{1\over3} \mu_{Fe(III)}$.  For a particle surface of fractal
dimension (as predicted by models in which the ferritin core is
formed by diffusion limited aggregation\cite{frankel}), the random
walk over the surface can give any power law from 1/3 to 1/2.  Lastly,
if all the ions (as opposed to merely the surface ones) are randomly
distributed between the sublattices, then $\mu$ =
$n^{1\over2}\mu_{Fe(III)}$.  This can occur, for example, if there is a
non-stochiometric replacement of some magnetic ions with non magnetic
ions.  This prediction (which has no free parameters) is plotted in
Fig.3 as a dashed line.  The agreement between this last prediction
and the data might be strong evidence for the random population of the
sublattices throughout the volume of the particles.  We note, however,
that for particles with 100 to 3000 magnetic ions the
discreteness of the lattice, combined with any surface roughness, means
that a disproportionately large number of sites will be located on the
surface.  Thus, it is not possible to determine whether the sublattice
non compensation is a volume effect or a surface effect.  That the
power law is clearly much less than 1 is strong evidence that the
excess moment does not result from canting of the sublattices.
Measurements made on fully loaded natural ferritin (n = 4500) and
partially loaded natural ferritin (n = 2000) are in good agreement
with a 1/2 power law.\cite{mohie,applic:Kil95,makhlouf}

We can now compare the measured values of the excess spin and the MQC
resonance frequency $\nu_{MQC}$ as a functions of $n$.  An
antiferromagnet strongly coupled to an uncompensated moment can be
described by the effective action\cite{model:Loss92,Chio97}

\begin{eqnarray}
S_E&=&V\int\!\!d\tau\Bigl\{
{\chi_\perp\over2\gamma^2}(\dot{\theta}^2+\dot{\phi}^2\sin^2\theta)
+K_y\sin^2\theta\sin^2\phi\nonumber\\
&&\qquad{}+K_z\cos^2\theta\Bigr\}+ i\hbar
S\int\!\!d\tau\,\dot{\phi}(1-\cos\theta){,}
\end{eqnarray}
where $\theta$ and $\phi$ are the spherical coordinates of the N\'eel
vector, $V$ is the volume of the grain, $K_z\geq K_y>0$ its magnetic
anisotropies, $\gamma=2\mu_B/\hbar$, and $S$ is the magnitude of the
excess spin. Let us define $S_{\rm AFM}=\hbar
V\sqrt{2K_y\chi_\perp}/\mu_B$, the instanton action one would obtain
for an antiferromagnet without an uncompensated moment.

We will use instanton techniques to calculate the tunnel splitting.
It has been shown\cite{Chio97} that in the regime $K_y\ll K_z$
instanton solutions have an approximate frequency
\be
\omega_{\rm Ferri}={2\lambda V\over\hbar S}\sqrt{K_yK_z}{,}
\ee
with
\be
\lambda=\left( 1+{1\over4}{K_z\over K_y}\left(S_{\rm AFM}\over\hbar
S\right)^2
\right)^{-1/2}{.}
\ee
and an action
\begin{eqnarray}
S_{\rm Ferri}&=&{2\hbar S\over\lambda}\sqrt{K_y/K_z}\biggl\{1+
\hbox{$\textstyle{1\over3}$}\lambda^2{K_y\over K_z}+\nonumber\\
&&\qquad\hbox{$\textstyle{1\over8}$}\lambda^4\Bigl({S_{\rm AFM}\over\hbar
S}\Bigr)^2\biggl(1+\delta\Bigl(\sqrt{K_y\over K_z}\lambda\Bigr)
\biggr)\biggr\}{.}
\end{eqnarray}
where
\be
\delta(x)={1\over x^3}\Bigl(\sqrt{1+x^2}\,\hbox{arcsinh}(x)-x-x^3/3\Bigr){.}
\ee
The tunnel splitting is then given by
\be
\Delta_0=8\hbar\omega_{\rm Ferri}\sqrt{S_{\rm Ferri}\over2\pi\hbar}
\left|\cos(\pi S)\right|
e^{-S_{\rm Ferri}/\hbar}{,}
\label{splitting}
\ee
and the crossover temperature to the quantum regime by
\be
k_BT^*=K_yV\hbar/S_{\rm Ferri}{.}
\label{Tcross}
\ee

We suppose $\chi_\perp$ is independent of loading and use the simple
estimate\cite{intro:Barb90,applic:Bar85}
$\chi_\perp=\mu_B^2N/k_BT_NV$.  The volume of the magnetic core of a
fully loaded grain is roughly that of a sphere with a diameter of
$7.5\,{\rm nm}$, and contains $4500$ ions.  $T_N=240\,\unite{K}$
in ferritin, which gives $\chi_\perp\approx5\cdot10^{-5}\,
{\unite{emu}/\unite{G}\,\unite{cm}^3}$.

The crossover temperature to the quantum regime has only been
measured\cite{intro:Awsch92} in fully loaded grains, where
$T^*=200\,{\rm mK}$.  In the absence of experimental data we will
assume $T^*$ is also independent of loading.

Armed with these values for $\chi_\perp$ and $T^*$, we are left with 
two unknowns at each loading, the anisotropies $K_y$ and $K_z$.  We 
can deduce their values from Eqs.\ (\ref{splitting}) and 
(\ref{Tcross}), using the experimental values for $\mu$ and 
$\nu_{MQC}$ listed in Table \ref{dependent}.  We see that the 
hypothesis that $K_y\ll K_z$ is at least self-consistent, and that  
the anisotropies are not strongly size dependent.  The only exception is the 
value obtained for $K_z$ in fully loaded grains.  We note that the 
measurement of $\mu$ in fully loaded natural ferritin\cite{applic:Kil95} 
lies somewhat below the power law shown in Fig.3, possibly as the 
result of slightly different sample preparation, interparticle effects 
(the samples were not diluted), or differences in the surface of a 
grain completely filling the spherical protein shell.  The value of 
$K_y$ increases somewhat for smaller particles, consistent with the 
trend in blocking temperature as a function of $n$,\cite{JAP:Gider96} 
though it is somewhat smaller than typical for 
antiferromagnetic particles.  We are not aware of any other 
measurements of the transverse anisotropy $K_z$ in such systems.

In conclusion, we have measured the excess moment of the
antiferromagnetic protein ferritin as a function of the number of
magnetic ions per protein.  Using diluted samples to ensure the
absence of interparticle interactions, we find an approximately square
root dependence of the excess moment upon particle size, in agreement
with a simple model (which has no free parameters) of imperfect
sublattice compensation.  We use this result to compare recent
theoretical work on the effect of an excess moment to earlier MQC and
blocking temperature measurements in the same samples.

We are grateful to Steven Mann and Trevor Douglas formerly of the
University of Bath, for providing the artificial ferritin samples.
This work was supported by the AFOSR \# F49620-99-1-0033.

\begin{figure}

\centerline{\epsfig{file=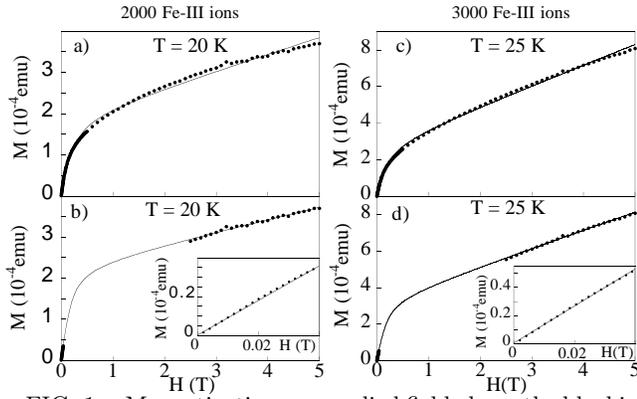, width = 1.0\linewidth}}

\caption{ Magnetization vs. applied field above the
blocking temperature for samples with particle size $n=2000$ Fe(III)
ions (a,b) and $n=3000$ (c,d).  The solid line is a fit to the
form of Eq.(\ref{modelM}), where in (b) and (d), only the low- and high-field
data are fit, as described in the text. The insets in (b) and (d) are
magnifications of the low-field data and fit.}

\label{M(H)}

\end{figure}

\begin{figure}

\centerline{\epsfig{file=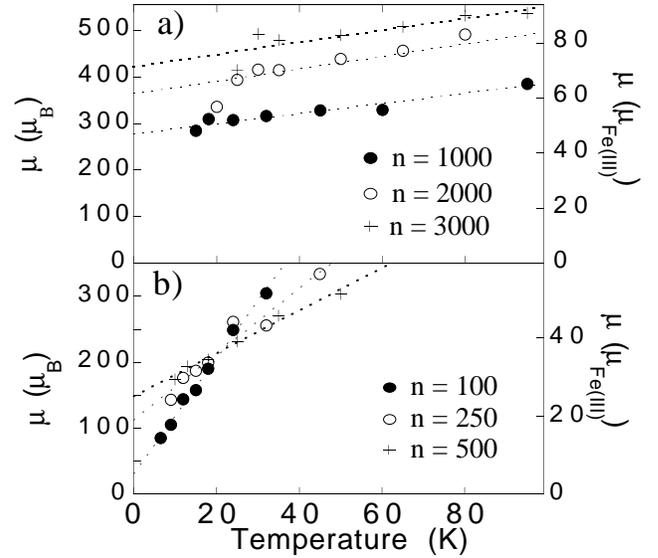, width = 1.0\linewidth}}

\caption{The excess moment $\mu$ (in units of $\mu_{B}$ and the ionic 
moment $\mu_{Fe(III)}$) of ferritin cores each with $n$ Fe(III) ions 
as a function of temperature from $T_B$ to roughly 4$T_B$ for each 
sample.  The dashed lines are linear fits to, in (a) the larger cores, 
and in (b) the smaller cores. The temperature scale is the same in 
both a) and b).}

\label{mu(T)}

\end{figure}

\begin{figure}

\centerline{\epsfig{file=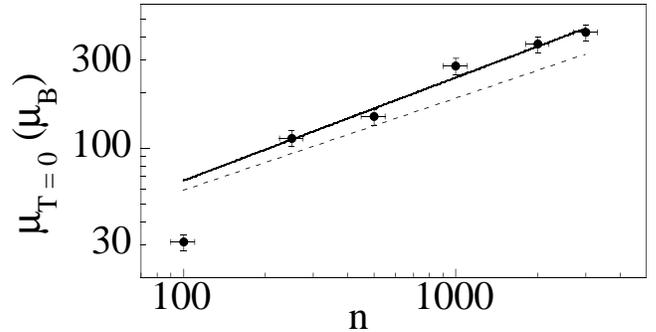, width = 1.0\linewidth}}

\caption{The T=0 excess moment in units of
$\mu_B$ as a function of particle size n.  The solid line is a power law
fit giving an exponent of 0.56.  The dashed line is the no free
parameter prediction of the $\mu = n^{1\over2}\mu_{Fe(III)}$ model
described in the text.}

\label{mu(n)}

\end{figure}

\begin{table}

\begin{tabular}{lllll}
$n$&$\mu$&$\nu_{MQC}$&$K_y$&$K_z$\\
&[$\mu_{B}$]&[$\rm Hz$]&[${\rm erg}/{\rm cm}^3$]&[${\rm erg}/{\rm cm}^3$]\\
\hline
$100$&$31$&&&\\
$250$&$113$&&&\\
$500$&$148$&&&\\
$1000$&$278$&$1.6\cdot10^8{}^\ast$&$3.8\cdot10^{3}$&$6.1\cdot10^{6}$\\
$2000$&$401$&$7.8\cdot10^7{}^\ast$&$2.1\cdot10^{3}$&$7.7\cdot10^{6}$\\
$3000$&$423$&$5.6\cdot10^6{}^\ast$&$1.9\cdot10^{3}$&$4.5\cdot10^{6}$\\
$4500$&$316^\natural$&$9.40\cdot10^5{}^\sharp$&$1.5\cdot10^{3}$&$1.9\cdot10^{6}$\\
\hline
\multicolumn{5}{l}{${}^\ast$ MQC measurements in artificial ferritin, ref.\
\onlinecite{intro:Gid95}}\\
\multicolumn{5}{l}{${}^\sharp$ MQC measurements in natural ferritin, ref.\
\onlinecite{intro:Awsch92}}\\
\multicolumn{5}{l}{${}^\natural$ Ref.\ \onlinecite{applic:Kil95}}
\end{tabular}

\caption{Excess moment $\mu$, MQC resonance frequency $\nu_{MQC}$, and
anisotropy energies $K_{y}$ and  $K_{z}$ for ferritin cores
containing $n$ Fe(III) ions.}
\label{dependent}

\end{table}

\end{document}